\documentclass[aps,prl,showpacs,twocolumn]{revtex4}
\usepackage[usenames]{color}
\newcommand{\comment}[1]{}
\usepackage{graphicx}
\usepackage{amsmath}
\usepackage{epsfig}
\newcommand{\bra}[1]{\langle {#1} |}

\newcommand{\expect}[1]{\langle {#1} \rangle}
\newcommand{\ketn}[1]{ {#1} \rangle}

\begin{document}

\title{Vortex lattice transitions in cyclic spinor condensates}

\date{\today}
\author{Ryan Barnett,$^1$ Subroto Mukerjee,$^{2,3}$ and Joel E. Moore$^{2,3}$}
\affiliation{$^{1}$Department of Physics, California Institute
of Technology, MC 114-36, Pasadena, California 91125, USA}
\affiliation{$^2$Department of Physics, University of
California, Berkeley, CA 94720}
\affiliation{$^3$Materials Sciences Division,
Lawrence Berkeley National Laboratory, Berkeley, CA 94720}

\begin{abstract}
We study the energetics of vortices and vortex lattices produced
by rotation in the cyclic phase of $F=2$ spinor Bose condensates.  In addition
to the familiar triangular lattice predicted by Tkachenko for $^4$He, many more complex
lattices appear in this system as a result of the spin degree of freedom. In
particular, we predict a magnetic-field-driven transition from a
triangular lattice to a honeycomb lattice.  Other transitions and
lattice geometries are driven at constant field by changes in the
temperature-dependent ratio of charge and spin stiffnesses, including
a transition through an aperiodic vortex structure.
\end{abstract}

\pacs{03.75 Mn}
\maketitle

One of the many remarkable properties of superfluids is the
appearance of vortex lattices in rotated
systems~\cite{tkachenko66, yarmchuk79}.  These lattices are
periodic arrangements of vortices that allow the superflow outside
the vortex cores to remain irrotational and are analogous to the
mixed state of type-II superconductors in a magnetic field.  Bose
condensates of atoms with nonzero integer spin~\cite{stenger98,ho98,ohmi98},
referred to as ``spinor Bose condensates'', combine spin
and superfluid ordering in different ways depending on the
spin and the interatomic interaction.  These condensates, and the vortices and other
topological defects that they allow, have been actively studied in
recent years.

Since the physics of individual defects in spinor Bose condensates is now
understood for the most experimentally relevant cases with total spin
$F\leq3$,~\cite{zhou01, makela06a, mukerjee06, semenoff07, yip07, barnett07},
a natural next step is to
understand physical situations controlled by the collective physics
of many defects.  Two examples are the vortex lattice in a rotated condensate
and the superfluid transition in a two-dimensional condensate.  In
general, the lowest-energy vortex defects of spinor condensates have both
superfluid and spin character. Although external rotation of the
condensate couples only to the superfluid part, the mixed nature of the
vortices means that the interaction between the spin parts of different
vortices is also important in determining the vortex lattice.

This letter uses a general approach to vortex
lattice phases in spinor condensates, including the
quadratic Zeeman anisotropy normally present in experimental
systems, to show that the cyclic phase of an
$F=2$ spinor condensate undergoes an unusual vortex lattice
transition in a weak applied magnetic field.  This transition
allows collective physics resulting from the nontrivial spin
configuration of vortices to be imaged using spin-insensitive
measurements.  The comparison of energies of different lattices
uses an Ewald summation trick that exactly reproduces previous
results obtained for simpler lattices using elliptic
functions~\cite{tkachenko66}.  More generally, the methods of this
letter allow the energy of any periodic arrangement of vortices in
a spin-anisotropic spinor condensate to be rapidly calculated.
We also show that under some conditions there is a strictly
aperiodic vortex structure rather than a true lattice.

Dilute $F=2$ bosons interact via the potential $V(|{\bf r}_1-{\bf
r}_2|)=\delta({\bf r}_1 - {\bf r}_2) (g_0 P_0 + g_2 P_2 + g_4
P_4)$, where $P_F$ projects into the total-spin $F$ state and $g_F
= 4 \pi \hbar^2 a_F / M$ determines $g_F$ given $a_F$, the
scattering length in the spin-$F$ channel.
This two-body potential gives the interaction
Hamiltonian~\cite{ciobanu00, ueda00}
\begin{equation}{\cal H}_{\rm int} = \int d {\bf r} :
{\alpha \over 2} (\psi^\dagger \psi)^2 + {\beta \over 2}
|\psi^\dagger {\bf F} \psi|^2 + {\tau \over 2} |\psi^\dagger
\psi_t|^2 :,
\end{equation}
with $\psi$ a five-component vector field whose component
$\psi_m({\bf r})$ destroys a boson at point ${\bf r}$ with $F_z =
m$, $m =-2, \ldots, +2$, and ${\bf F}$ denoting the spin-2
representation of the $SU(2)$ generators.  $\psi_t$ is the
time-reversal conjugate of $\psi$: $\psi_{tm} = (-1)^m
\psi^\dagger_m$.  The parameters in this Hamiltonian are
determined by $g_0, g_2, g_4$ via $\alpha = (3 g_4 + 4 g_2)/7$,
$\beta = -(g_2 - g_4)/7$, $\tau = {1 \over 5} (g_0 - g_4) - {2
\over 7} (g_2 - g_4).$ To $H_{\rm int}$ must be added the one-body
Hamiltonian for an isotropic and spatially uniform condensate
 \begin{equation} 
{\cal H}_0 = \int d
{\bf r} \; \frac{\hbar^2}{2M} \nabla \psi^{\dagger} \cdot \nabla \psi
- \mu \psi^\dagger \psi,
\end{equation}
where $\mu$, the chemical potential. Minimizing this Hamiltonian
over single-particle condensates leads to three phases:
ferromagnetic, antiferromagnetic, and cyclic. The cyclic phase
that will be the focus of our work occurs when $\beta, \tau
>0$ and is expected to be realized in a condensate of $^{85}$Rb
atoms \cite{ciobanu00}.  The spinor structure of this state,
having the symmetry of the tetrahedron, results
in a nonabelian homotopy group which has been pursued in
the liquid physics community for several years.

In all existing experiments, an important effect even at the
single-particle level is the existence of anisotropy in spin space
resulting from the magnetic fields used as part of the trapping
process. Including the hyperfine interaction, the bosons we
consider interact with the external magnetic field as ${\cal H}_z
= \Gamma {\bf I} \cdot {\bf S} - 2 \mu_B B_z$ where $\Gamma$ is
the magnitude of the hyperfine interaction, ${\bf I}$ is the
nuclear spin, ${\bf S}$ is the electronic spin, $\mu_B$ is the
Bohr magneton, and $B$ is the magnitude of the magnetic field
taken to point in the $z$-direction. Within the manifold of
spin-two states, a Hamiltonian which reproduces the correct
energies up to a constant is given by ${\cal H}_z=\sqrt{\Gamma^2
+(\mu_B B)^2 + \Gamma \mu_B B F_z}$ \cite{briet31}. 
This Hamiltonian can be
expanded in powers of $F_z$.  Since the relaxation time of the
total magnetization is typically longer than the condensate
lifetime, the linear term can be neglected.  Particular attention
has been paid to the next term which gives rise to the quadratic
Zeeman effect \cite{stenger98}. However, due to the high symmetry
of the cyclic state, this quadratic term alone is not enough to
select its orientation.  For this case, one therefore must
consider the cubic term which is at next order.

\begin{figure}
\includegraphics[width=3.5in]{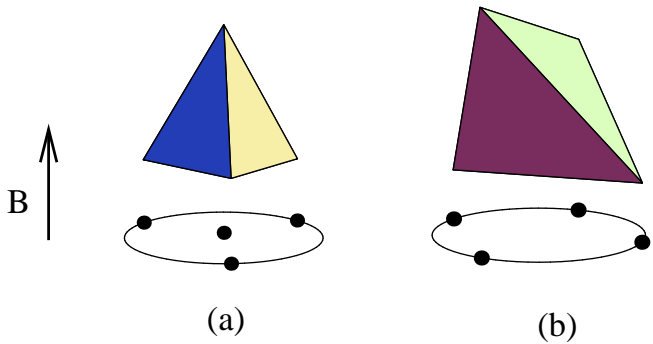}
\caption{Orientations of the cyclic state in an external magnetic
field which breaks the spin rotational symmetry.  Upon increasing
the magnetic field, the spinor will undergo a transition from state
(a) to state (b).}
\label{Fig:shapes}
\end{figure}

To determine the spin states in the presence of such a magnetic
field, one must also consider the spin exchange interaction energy
of the condensate per particle which is $E_s=\frac{1}{2} n \beta
\expect{\bf F} \cdot \expect{\bf F} +\frac{1}{2} n \tau
|\bra{\chi}\ketn{\chi_t}|^2$, where $n$ is the condensate density. Since
the total spin is assumed to be conserved in experiments, we can neglect the first
term in this expression. Minimizing $E_s+ E_z$ (where
$E_z=\expect{{\cal H}_z}$) over possible spinor states we find the
following: At small magnetic fields, the spinor
$\chi_{t1}=\left(\sqrt{1/3},0,0,\sqrt{2/3},0\right)^T$ (up to any
rotation about the $z$-axis) is selected.  In the classification
scheme described in \cite{barnett06}, this state is represented by
a tetrahedron with one of its faces parallel to the $xy$ plane.
Upon increasing the magnetic field  there is a transition at
$\mu_B B_c = n \tau/16$ to the spin orientation
$\chi_{t2}=\left(\sin(\theta)/\sqrt{2},0,\cos(\theta),0,
-\sin(\theta)/\sqrt{2}\right)^T$, where $\theta$ changes
continuously with increasing magnetization. This spinor has the
symmetries of a distorted tetrahedron with one of its edges
parallel to the $xy$ plane (when $\theta=\pi/4$ it has the
symmetries of the regular tetrahedron). These two types of
orientations are summarized in Fig.~\ref{Fig:shapes}. 
The magnitude
of the critical field is of the order of ambient fields in current
experiments \cite{sadler06} but smaller fields
can in principle be simulated by optical means \cite{gerbier06}.

Having identified the different types of tetrahedral states that
are stabilized in an applied magnetic field, we now discuss the
effects of rotation in addition to the applied field. The rotation
couples to the phase of the condensate and has the effect of
producing point vortices in two dimensions or lines of vortices in
three dimensions.  A vortex is a special type of configuration in an ordered phase
that breaks a continuous symmetry: sufficiently far from a
``core'' region (linear in 3D or point-like in 2D) in which the
order is destroyed. The configuration is locally in an ordered
state, but cannot be smoothly deformed to the uniform
configuration.

We find in general that vortices form a two-dimensional lattice whose
properties depend on the nature of the constituent vortices and
the interactions between them. For simplicity, it is assumed that
the magnetic field and the axis of rotation are in the same
direction. Owing to the $\pi/3$ spin rotation symmetry of the
state (a) that is stabilized at fields $B < B_c$, its vortices are
of three types: $(n,m)$, $(n-1/3,m+1/3)$ and
$(n+1/3,m-1/3)$, where $n$ and $m$ are integers and the first argument
inside the parentheses is the winding number of the phase while
the second is that of the spin. The vortex lattice that is formed
has a net nonzero winding number for the phase and zero winding
number for the spin.

The energetics, in addition to the above constraints on the
winding numbers, will depend on the stiffnesses $K_c$ and $K_s$ of
the condensate corresponding respectively to the charge (phase)
and the spin.  (The expected behavior of these stiffnesses will be discussed in
closing.)  The interaction energy of two vortices $(x_1,y_1)$
and $(x_2,y_2)$ separated by a distance $r$ in the state (a) is
given by
\begin{equation}
E = 2\pi K_cx_1x_2\log(\xi/r)+\pi K_sy_1y_2\log(\xi/r),
\label{Eq:vorten}
\end{equation}
where $\xi$ is the typical radius of a vortex. For $K_s > K_c$, it
is energetically favorable to produce only vortices of the type
$(1,0)$. However, for $K_s/K_c < 1$, the $(1,0)$ vortex breaks up
into $(2/3,1/3)$ and $(1/3,-1/3)$ vortices as can be seen by
putting the appropriate values of the winding numbers into
Eqn.~\ref{Eq:vorten}. Once again, the energetics are subject to
the constraints of rotation mentioned in the previous paragraph.
For values of $K_s/K_c < 1/4$, each $(2/3,1/3)$ vortex breaks up
into $(1/3,2/3)$ and $(1/3,-1/3)$ vortices. Thus, in this regime,
there are only $(1/3,2/3)$ and $(1/3,-1/3)$ vortices with twice as
many of the latter as the former. Since, the nature of the
vortices is determined by the ratio $K_s/K_c$, so too is the
lattice they form as will be described later. For $B > B_c$, the
state (b) is stabilized. This state can be shown to have vortices
only of the type $(n,m)$. When subjected to a rotation only
$(1,0)$ vortices will be produced like at low fields with $K_s >
K_c$.

Thus, to summarize, the following kinds of vortices are produced
by rotation: 1) $(1,0)$ vortices for $K_s
> K_c$ or $B > B_c$, 2) an equal number of $(2/3,1/3)$ and
$(1/3,-1/3)$ vortices for $1/4 <K_s/K_c < 1$ and $B < B_c$ and 3)
twice as many $(1/3,-1/3)$ vortices as $(1/3,2/3)$ for $0 <
K_s/K_c < 1/4$ and $B < B_c$. In each case, the density of
vortices is determined by the angular velocity of the rotation.

Having determined the types of vortices that are produced at the
various values of magnetic field and stiffnesses, we now
evaluate the energies of vortex lattices. Due to
the long-ranged nature of the logarithmic interactions, the energy
of a vortex lattice is difficult to evaluate directly. Thus, we
develop a method that is similar to the Ewald summation technique
for the cohesive energy of
three-dimensional ionic crystals \cite{ewald21,tosi64}. For
simplicity, we use a scalar condensate to demonstrate the
technique; the generalization to spinor condensates is
straightforward and will be given presently. The energy (in units of
the stiffness) of a single vortex taken to be at the origin is
given by
\begin{equation}
\phi(0)= \sum_{{\bf R} \ne 0} \log\left(\frac{\xi}{\bf R}\right) -
\int d^2 r \rho_0 \log\left(\frac{\xi}{r}\right),
\label{Eq:latsum}
\end{equation}
where ${\bf R}$ are the lattice vectors and $\xi$, the size of the
vortex. The second term is due to a uniform negative background
charge of density $\rho_0$ which arises from the fact we are
working in a rotating frame of reference. Note that each of these
terms diverges individually but their difference does not. The
Ewald trick is to add and subtract a normalized Gaussian function
$\frac{1}{\pi \sigma^2} e^{-r^2/\sigma^2}$ from each point charge,
where $\sigma$ is a screening length. For instance, the potential
of a point charge at the origin screened by such a Gaussian
function is
$
\phi(r) = \frac{1}{2} {\rm Ei}(r^2/\sigma^2),
$
where ${\rm Ei}(x) = \int_x^{\infty} dt e^{-t}/t$ is the
exponential integral.  Note that this will decay exponentially fast at
large distances.  In three dimensions
the corresponding potential is $\phi(r) = (1-{\rm erf}(r))/r$ where erf
is the error function.

Proceeding along these lines,
the resulting potential corresponding to Eq.~\ref{Eq:latsum} is
\begin{align}
\phi(0) =& \frac{1}{2} \sum_{{\bf R} \ne 0}
{\rm Ei}\left(\frac{R^2}{\sigma^2}\right)
+ \sum_{{\bf G} \ne 0} \rho_0 \frac{2\pi}{G^2} e^{-\frac{\sigma^2}{4}G^2}
\notag
\\
&- \log\left(\frac{\xi}{\sigma}\right)-\frac{\gamma}{2} - \rho_0
\frac{\pi}{2}\sigma^2,
\label{Eq:ewald1}
\end{align}
where ${\bf G}$ are the reciprocal lattice vectors and  $\gamma$
is the Euler-Mascheroni constant. The first term comes from the
density of point charges screened by the Gaussian function while
second term comes from difference of the charge density of the
Gaussian functions and the uniform charge density.  The term $-
\log\left(\frac{\xi}{\sigma}\right)-\frac{\gamma}{2} $ is obtained
after subtracting off the additional Gaussian function at the
origin where we omit the charge.  Finally, the last term is to
make the average of the screened potential zero \cite{tosi64}.
On the other hand, the potential of a test charge away by ${\bf d}$
from
the origin is
\begin{align}
\phi({\bf d}) = \frac{1}{2} \sum_{{\bf R} } 
{\rm Ei}\left(\frac{|{\bf R}-{\bf d}|^2}{\sigma^2}\right)
&+ \sum_{{\bf G} \ne 0} \rho_0 \frac{2\pi}{G^2} e^{-\frac{\sigma^2}{4}G^2}
e^{i{\bf G} \cdot {\bf d}}
\notag
\\ &- \rho_0 \frac{\pi}{2}\sigma^2
\label{Eq:ewald2}
\end{align}
The best check of this procedure is to see if the sum is
independent of the parameter $\sigma$.  The two sums in real and
reciprocal space in Eqns.~\ref{Eq:ewald1} and \ref{Eq:ewald2} both
converge exponentially fast. In this way, the energy per vortex of the square
lattice is found to be $\phi_s(0) = -\log\left(
\frac{\xi}{a}\right)-1.3105$ (where $a$ is the lattice constant)
while that of the triangular lattice at the same density is found
to be $\phi_t(0) = -\log\left( \frac{\xi}{a}\right)-1.3211$.  Both
are in precise agreement with the results obtained by integrating
over the full spatial flow pattern~\cite{tkachenko66}.  For vortices in spinor condensates, which contain windings of phase and spin, the above procedure is applied
individually to each sector with the Ewald sums being weighted by
the corresponding stiffnesses. The main advantage of the Ewald
technique is that it can be generalized to treat complicated unit
cells with an arbitrary number of vortices in them in a
straightforward and numerically efficient manner.

\begin{figure}
\includegraphics[width=3.5in]{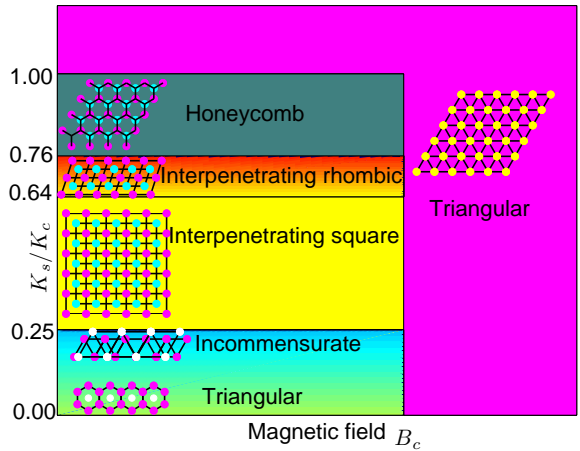}
\caption{A schematic depiction of the different types of vortex
lattices obtained at different values of $K_s/K_c$ and magnetic
field. The color code for the vortices is: (1,0) yellow, (2/3,1/3)
cyan, (1/3,-1/3) magenta and (1/3, -2/3) white.}
\label{Fig:phasediag1}
\end{figure}

Let us first consider the case $B>B_c$. As noted earlier, the
vortices produced by the rotation are of the type $(1,0)$. These
form the usual triangular lattice for all values of
$K_s/K_c$. For $B<B_c$, the fractional winding numbers of the
fundamental vortices give rise to more interesting possibilities.
For $1/4<K_s/K_c<1$, the lattice is bipartite with equal numbers
of $(2/3,1/3)$ and $(1/3,-1/3)$ vortices. We use the Ewald
summation technique to numerically evaluate the energy of the
lattice assuming the same parallelogram unit cell for both
sublattices and an arbitrary displacement between them. We then
perform a minimization of the energy over these parameters to
identify the lattice that is produced at different values of
$K_s/K_c$. The sequence of lattices is described in
Fig.~\ref{Fig:phasediag1}. At exactly $K_s/K_c=1$, the two
sublattices do not interact with each other and each is a
triangular lattice. As soon as $K_s/K_c$ is lowered and the two
begin to interact, the honeycomb lattice is stabilized and remains
so till $K_s/K_c=0.76$. Below this value, the vortices of one type
move to the centers of the rhombic unit cells formed by the other
type which we term an interpenetrating rhombic lattice. The
internal angle of the rhombus changes continuously with $K_s/K_c$
from $\pi/3$ at $K_s/K_c=0.76$ to $\pi/2$ at $K_s/K_c=0.64$. The
interpenetrating square lattice thus obtained at $K_s/K_c=0.64$ is
stable down to $K_s/K_c=1/4$.  This sequence of lattices is the same as
obtained for rotating two-component condensates in the quantum
Hall regime~\cite{muellerho02}, or equivalently the $F=1$ polar
condensate,
but the values where the transitions occur are different for $F=2$.

For $K_s/K_c<1/4$, a lattice with $(1/3,2/3)$ and $(1/3,-1/3)$
vortices is obtained with twice as many of the latter as the
former. Exactly at $K_s/K_c=1/4$, the two sublattices do not
interact and each is a triangular lattice. The sublattice of the
$(1/3,2/3)$ vortices has a unit cell of length $\sqrt{2}$ times
that of the $(1/3,-1/3)$ vortices.  These two lattices are {\it incommensurate}
for any rotation angle between them, which follows from showing that
the nonzero squared lengths of lattice vectors in one lattice are disjoint from
those in the other lattice.
This incommensurability implies that the energy of interaction
between the two lattices can be calculated using the Ewald technique by
averaging over all displacement vectors instead of specific lattice points,
and the result is zero. In the other limit, $K_s/K_c \rightarrow 0$,
the interaction between all pairs of vortices is
identical and a triangular lattice is obtained. While there are
several way to distribute the two kinds of vortices in such a
lattice, the lattice where the $(1/3,-1/3)$ vortices form a
honeycomb lattice while the $(1/3,2/3)$ vortices are at the
centers of each hexagon is the most symmetric one with three
vortices per unit cell.  The behavior between the incommensurate
structure at $K_s/K_c = 1/4$ and this specific triangular structure
as $K_s/K_c \rightarrow 0$ is difficult to determine reliably by our
technique, since given the existence of the incommensurate structure, there is
no justification for a numerical search over unit cells with a finite
number of basis vectors.

As demonstrated above, transitions between different vortex
lattices can be tuned by a magnetic field $B$ or the ratio
$K_s/K_c$. While the field $B$ can be applied directly or its
effect simulated through optical techniques in
experiments~\cite{gerbier06}, the ratio $K_s/K_c$ is more difficult
to manipulate directly. In spinor condensates at low temperatures,
this ratio is typically close to 1 but is renormalized by
quantum and thermal fluctuations.  Increasing temperature
acts to reduce $K_s$ more rapidly than $K_c$, because the soft spin modes
that are excited at finite temperature have a larger phase space than
the phase modes (assuming that the quadratic Zeeman term can be
neglected).  Both a nonlinear-sigma-model analysis and a study of
Bogoliubov-like excitations suggest that under normal experimental
conditions at nonzero temperature, $K_s$ is slightly less than $K_c$
so that the magnetic transition will be observable.  The best possibility
to observe evolving vortex structure as $K_s/K_c$ is further reduced
is to raise the temperature very close to $T_c$ of the condensate:
if the magnetic order is lost before the superfluid order, as allowed by
Landau theory, this ratio will rapidly decrease to zero in a narrow temperature
range.

To conclude, we have shown using the Ewald summation technique that
different types of vortex lattices can be produced in cyclic
condensates as functions of magnetic field and the ratio of the
charge and the spin stiffnesses. In particular, there is a
magnetic-field-driven transition from a triangular to a honeycomb
lattice. In the low-field limit, there are both abrupt transitions and
continuous families of lattices as functions of the ratio of the
stiffnesses, including the appearance of an incommensurate
structure at one value.

The authors would like to thank D. A. Huse, M. Lucianovic, 
O. Motrunich, D. Podolsky, D.
Stamper-Kurn, M. Vengalattore, and A. Vishwanath for useful
discussions. RB was supported by the Sherman Fairchild Foundation,
SM by DOE/LBNL, and JEM by NSF DMR-0238760.

\end{document}